\documentclass{article}

\usepackage{PRIMEarxiv}

\usepackage[utf8]{inputenc}
\usepackage[T1]{fontenc}  
\usepackage{hyperref}      
\usepackage{url}           
\usepackage{booktabs}       
\usepackage{amsfonts}    
\usepackage{nicefrac}     
\usepackage{microtype}    
\usepackage{lipsum}
\usepackage{fancyhdr}    
\usepackage{graphicx} 
\usepackage{kantlipsum}
\usepackage{statmath}
\usepackage{bm}
\usepackage{caption}
\usepackage{subcaption}
\usepackage{multirow}
\usepackage{xcolor}

\title{Directional Gaussian spatial processes for South African wind data}

\author{
  Jacobus S. Blom \\
  Department of Statistics, University of Pretoria, Pretoria, South Africa \\
   \And
  Priyanka Nagar$^*$ \\
  Department of Statistics and Actuarial Sciences, Stellenbosch University, Stellenbosch, South Africa\\
  \texttt{pnagar@sun.ac.za} \\
     \And
  Andri\"ette Bekker \\
  Department of Statistics, University of Pretoria, Pretoria, South Africa \\ Centre of Excellence in Mathematical and Statistical Sciences, Johannesburg, South Africa \\
}

\date{\today}

\begin{document}

\maketitle

\begin{abstract}
Accurate wind pattern modelling is crucial for various applications, including renewable energy, agriculture, and climate adaptation. In this paper, we introduce the wrapped Gaussian spatial process (WGSP), as well as the projected Gaussian spatial process (PGSP) custom-tailored for South Africa's intricate wind behaviour. Unlike conventional models struggling with the circular nature of wind direction, the WGSP and PGSP adeptly incorporate circular statistics to address this challenge. Leveraging historical data sourced from meteorological stations throughout South Africa, the WGSP and PGSP significantly increase predictive accuracy while capturing the nuanced spatial dependencies inherent to wind patterns. The superiority of the PGSP model in capturing the structural characteristics of the South African wind data is evident. As opposed to the PGSP, the WGSP model is computationally less demanding, allows for the use of less informative priors, and its parameters are more easily interpretable. The implications of this study are far-reaching, offering potential benefits ranging from the optimisation of renewable energy systems to the informed decision-making in agriculture and climate adaptation strategies. The WGSP and PGSP emerge as robust and invaluable tools, facilitating precise modelling of wind patterns within the dynamic context of South Africa.
\end{abstract}


\section{Introduction}\label{sec:intro}

The objectives, tactics, long-term aspirations, and growth trajectory pertaining to renewable energy under the framework of Sustainable Development Goal 7 (SDG-7) in the United Nations' 2030 Sustainable Development Goals\footnote{\url{https://sdgs.un.org/goals}[accessed 31 October 2023]} (SDGs) are designed to facilitate universal access to power, clean cooking fuels, and advanced technologies.  A concise overview of the latest findings and methodologies pertaining to the conversion of energy derived from renewable sources into usable forms is presented by \cite{trinh2023renewable}.  Over the past decade, there has been a notable growth in the proportion of the worldwide population that has obtained access to electricity, marking a significant milestone. However, it is worth noting that the number of individuals lacking access to electricity in Sub-Saharan Africa has experienced a concerning rise during the same period \footnote{\url{https://www.iea.org/reports/sdg7-data-and-projections/access-to-electricity}[Accessed 31 October 2023]} . South Africa must take measures toward the implementation of renewable energy initiatives in a global context where the popularity of fossil fuels is waning and climate action is viewed as an absolute necessity. Wind power could provide a remedy to South Africa's persistent energy shortages. Nevertheless, harnessing wind energy is a complex endeavour that requires a nuanced understanding of a variety of factors. The study of wind energy holds significant relevance in promoting the four key aspects of energy access, energy efficiency, renewable energy, and international collaboration, hence facilitating the advancement of Sustainable Development Goals. Therefore, modelling wind patterns is crucial in modern society for multiple reasons, including renewable energy, weather forecasting, air quality, and aviation.

Numerical models for weather forecasts require statistical post processing. Linear variables such as wind speed can be post processed in different ways as shown in \cite{lasinio2007statistical}, \cite{kalnay2002atmospheric} and \cite{wilks2006comparison}, whereas a circular (or angular variable) like wind direction cannot be post processed using standard methods (\cite{engel2007performance} and \cite{bao2010bias}). Bias correction and ensemble calibration techniques for determining the direction of wind are discussed in \cite{bao2010bias}. For the bias correction, \cite{bao2010bias} considered a circular-circular regression model as proposed in \cite{kato2008circular} and for the ensemble calibration a Bayesian model averaging with the von Mises distribution was considered. However, this study did not consider the spatial configuration in the data. The challenge is incorporating structured dependence into directional data.  Directional statistics have been developed for many years now starting as early as 1961, where the authors studied complex circular distributions underlying the theoretical framework (\cite{watson1961goodness}, \cite{stephens1963random} and \cite{kent1978limiting}). Various approaches to dealing with circular data, distribution theory and inference can be found in \cite{ley2017modern}, \cite{jupp2009directional} and \cite{mardia1972statistics}.  Previous studies conducted by \cite{rad2022enhancing} and \cite{arashi2020joint}  explore the feasibility of predicting wind direction in South Africa. Nevertheless, the inclusion of the spatial component in these studies was also lacking.

In the past, spatial models were employed to model wind patterns, but they had challenges with accounting for wind's nonlinear and complicated behaviour. Due to the spatial dependence structure that arises in wind data, a straightforward linear model cannot be used to model wind patterns, as discussed in \cite{jona2012spatial}. \cite{coles1998inference} proposed a wrapped Gaussian model for modelling wind directions. The approach assumed an unspecified covariance matrix and independent angular information, working in low dimensions. However, an extension to a spatial framework was briefly discussed. This extension was later introduced by \cite{casson1998extreme} where the circular variables were considered to be conditionally independent von Mises distributed. More recent, \cite{jona2012spatial} introduced a model to analyse wave direction data using a wrapped Gaussian spatial process (WGSP). The WGSP takes into account the spatial structure of directional variables with a potential for high dimensional multivariate observations which are driven by a spatial process. The methodology allows for the implementation of spatial prediction of the mean direction and concentration whilst also capturing the dependence structure. 

In this paper, we consider the WGSP and projected Gaussian spatial process (PGSP) for modelling wind patterns in South Africa. These models account for the highly complex dependence structure that arises in wind data as well as the periodic nature of directional data as developed by \cite{jona2012spatial}. (See  also \cite{ley2018applied}.) There are significant distinctions between the two approaches. The wrapping approach constructs a circular distribution that is similar (generally) to its real line counterpart. In other words, if the real line distribution is symmetric and unimodal then the wrapped distribution will have the same characteristics (\cite{jammalamadaka2001topics}). The projected Gaussian model, however, may result in differing characteristics from the real line counterpart. For example, the projected Gaussian model can be asymmetric and bimodal. The main justification for proposing these two techniques resides in that it is simple to introduce spatial dependence. The wrapping produces results that are relatively simple to interpret in terms of phenomenon behaviour, whereas the projection is extremely useful when interpretation is less critical and a highly flexible model is required as stated in \cite{mastrantonio2016spatio}.

The remainder of the paper follows as: Section \ref{sec:data} explores a South African wind data set to monitor the wind behaviour over the course of a day. Section \ref{sec:method} outlines the WGSP and the PGSP models. Section \ref{sec:results}  examines the behaviour of two distinct methodologies employed for evaluating the wind direction over multiple locations in South Africa. In Section \ref{sec:con}, we will delve into the last remarks and potential avenues for future research.\\

\section{Wind direction data in South Africa}\label{sec:data}

\begin{figure}[h]
	\centering
    \includegraphics[scale=0.4]{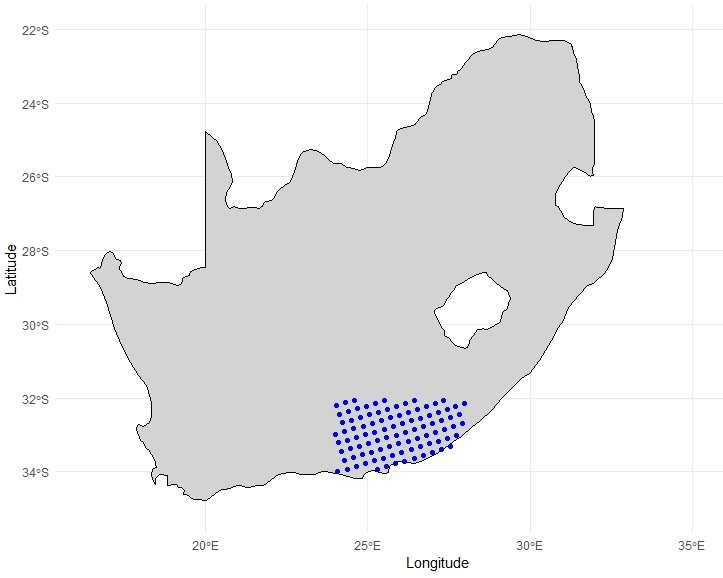}
	\caption{Map of South Africa with region under consideration indicated with dots.}
	\label{fig5}
\end{figure}

The data utilised was obtained from the Council for Scientific and Industrial Research (CSIR) database \footnote{\url{http://wasadata.csir.co.za/wasa1/WASAData} [Accessed July 2023]}. Data from $97$ locations which are relatively close to each other are considered for four different time periods (South African Standard Time (SAST)) on a particular day; 2012-12-31-05:00,  2012-12-31-11:00, 2012-12-31-17:00 and 2012-12-31-23:00. The region under consideration spans between: $32.054^\circ$ S, $24.009^\circ $ E and $33.992^\circ$ S, $27.99^\circ$ E, which gives an area of about $214.908$ $km$ $\times$ $370.723$ $km$ which is roughly $79671.338$ $km^2$. This region is illustrated in Figure \ref{fig5}. The data set included the wind direction in degrees, the longitude coordinate and the latitude coordinate.

The original wind direction is recorded in degrees, indicating the direction from which the wind originates. This is known as the meteorological wind direction (see  \cite{riha2020hyperpriorsensitivity}). Wind directions in degrees are converted to radians. Thus, a reading of $360^\circ$ or $0^\circ$, which is equal to $0$ rad or $2\pi$ rad, indicates wind coming from the north. Similarly, a reading of $90^\circ$, equivalent to $\pi/2$ rad, shows wind from the east, while $180^\circ$ or $\pi$ rad denotes wind from the south, and so on. The longitude and latitude coordinates are formatted in the Universal Transverse Mercator (UTM) format. Table \ref{tab:descriptives} provides the circular descriptive statistics of the wind direction over the entire region under consideration for the four different time periods. 

\begin{table}[h]
\centering
\caption{Circular descriptive statistics of the wind direction over the entire region under consideration for each time period.}
\label{tab:descriptives}
\begin{tabular}{lllll}
Time of day & mean direction & median direction & variance & standard deviation \\ \hline
05:00 & $0.62854$ & $0.55833$ &  $0.23251$ & $0.72750$ \\ \hline
11:00 & $0.51324$ & $0.52081$ & $0.10676$ & $0.47517$\\ \hline
17:00 & $0.14884$ & $0.11990$ &  $0.05948$ & $0.35019$ \\ \hline
23:00 & $0.20445$   & $0.18064$ & $0.07174$ &  $0.38585$ \\ \hline                 
\end{tabular}
\end{table}

\begin{figure}[h!]
     \centering
     \begin{subfigure}[b]{0.48\textwidth}
         \centering
         \includegraphics[width=\textwidth]{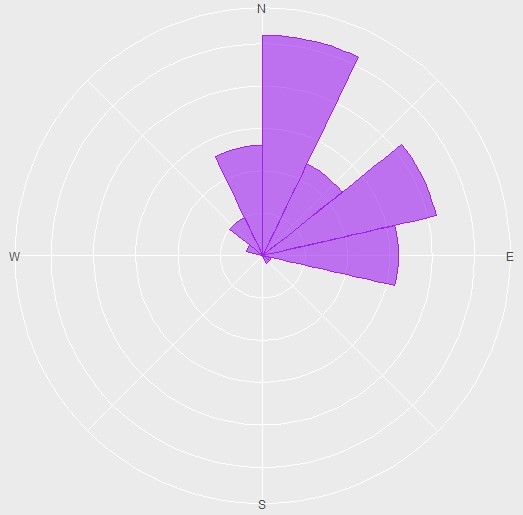}
         \caption{Rose diagram of wind direction over the entire region at 05:00.}
     \end{subfigure}
     \hfill
     \begin{subfigure}[b]{0.48\textwidth}
         \centering
         \includegraphics[width=\textwidth]{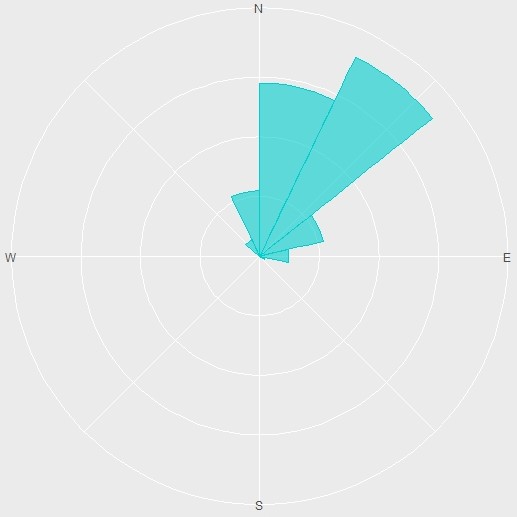}
         \caption{Rose diagram of wind direction over the entire region at 11:00.}
     \end{subfigure}
     ~
      \begin{subfigure}[b]{0.48\textwidth}
         \centering
         \includegraphics[width=\textwidth]{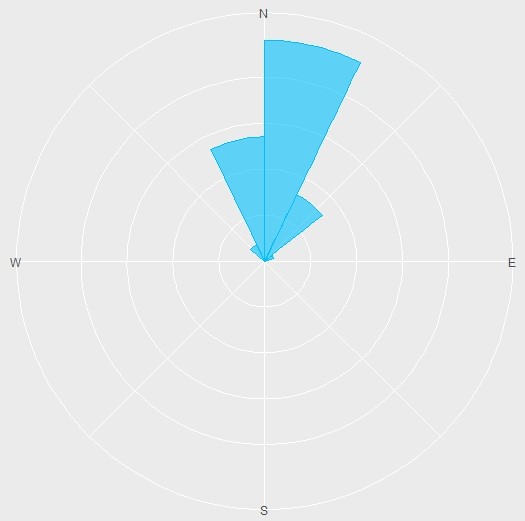}
         \caption{Rose diagram of wind direction over the entire region at 17:00.}
     \end{subfigure}
     \hfill
      \begin{subfigure}[b]{0.48\textwidth}
         \centering
         \includegraphics[width=\textwidth]{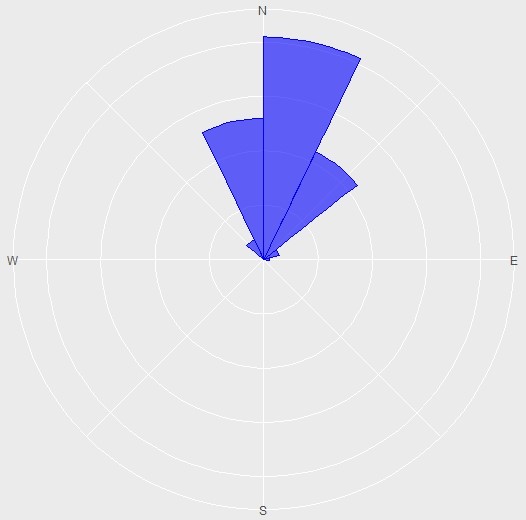}
         \caption{Rose diagram of wind direction over the entire region at 23:00.}
     \end{subfigure}
        \caption{Rose diagram of wind direction over the entire region for the four different time periods.}
        \label{fig4}
\end{figure}

The dominant wind direction for the region under consideration is Northerly and North-Easterly at 05:00, North-Easterly at 11:00, Northerly at 17:00 and 23:00 as shown in the rose diagrams presented in Figure \ref{fig4}. Based on the descriptive measures in Table \ref{tab:descriptives} and rose diagrams in Figure \ref{fig4}, it can be noted that the wind behaviour displays similar dominant directions for the two morning time periods (05:00 and 11:00) with a more North-Easterly pattern and the two evening time periods (17:00 and 23:00) with a Northerly pattern.

\section{Methodology}\label{sec:method}

\subsection{Wrapped Gaussian Spatial Process}

In the linear domain, suppose we define a multivariate distribution for $\textbf{Y}=(Y_1,Y_2,...,Y_p)$ with $\textbf{Y}\sim g(.)$, where $g(.)$ is a $p$-variate distribution on $\mathbb{R}^p$ indexed by $\omega$; a sensible choice for $g(.)$ would be a $p$-variate Gaussian distribution. Let $\textbf{K}= (K_1,K_2,...,K_p)$ be such that $\textbf{Y}=\textbf{X}+2\pi \textbf{K}$. Then $\textbf{X}=(X_1,X_2,...,X_p)$ is defined as a wrapped multivariate distribution induced from $\textbf{Y}$ with the transformation $\textbf{X}=\textbf{Y}$ mod $2\pi$. If the linear variable $\textbf{Y}$ is defined on $\mathbb{R}^p$ then the wrapped induced variable $\textbf{X}$ will also be defined on $\mathbb{R}^p$ as defined in \cite{jupp2009directional}. The wrapped Gaussian process will be fitted within a Bayesian framework using Markov Chain Monte Carlo (MCMC) methods, for further details the reader is referred to \cite{jona2012spatial}.

For the interpolation step, kriging will be used to make predictions at unobserved locations. Consider a Gaussian process (GP) in a spatial setting, we have locations $s_1,s_2,...,s_p$ where $s \in \mathbb{R}^d$ and $Y(s)$ is a GP with mean $\mu(s)$ and an exponential covariance function $\sigma^2 \rho(s-s';\phi)$ where $\phi$ is known as the decay parameter. We then have that $\bm{X}=(X(s_1),X(s_2),...,X(s_p))$ follows a wrapped Gaussian distribution with parameters $\bfmu = (\mu(s_1),...,\mu(s_p))$ and $\sigma^2 \textbf{R}(\phi)$ where $R(\phi)_{ij}=\rho(s_i - s_j ;\phi)$ as defined in \cite{jona2012spatial}. Suppose we have observations, $\textbf{X} = (X(s_1),X(s_2),..., X(s_p))$, and would like to predict a new value $X(s_0)$ at an unobserved location $s_0$. The point of departure follows similarly to a GP in the inline (linear) domain. The joint distribution for the linear observations $\textbf{Y} = (Y(s_1),Y(s_2),...,Y(s_p))$ along with the unobserved $Y(s_0)$ is given as:
	
	\begin{equation}\label{6}
		\begin{bmatrix}
			\textbf{Y}\\
			Y(s_0)
		\end{bmatrix} \sim \text{N}\left(\begin{bmatrix}
			\bfmu\\
			\mu(s_0)
		\end{bmatrix},\sigma^2 \begin{bmatrix}
			\textbf{R}_\mathbf{Y}(\phi) & \bfrho_{0,\mathbf{Y}}(\phi)\\
			\bfrho^{T}_{0,\mathbf{Y}}(\phi) & 1
		\end{bmatrix}\right),
	\end{equation}

 From (\ref{6}), the conditional distribution of of $Y(s_0)|\textbf{Y},\bftheta$ can be obtained. The wrapped Gaussian distribution of $X(s_0)|\textbf{X},\textbf{K},\bftheta$ and, thus, $E(e^{iX(s_0)}|\textbf{X},\textbf{K};\bftheta)$ can then easily be derived.  To obtain $E(e^{iX(s_0)}|\textbf{X},\textbf{K};\bftheta)$ it is necessary to marginalise over the distribution of $\textbf{K}|\textbf{X}, \bftheta$ which will require a $n$-fold sum over a multivariate discrete distribution which is problematic even when considering truncation. Thus, we consider a Bayesian framework, to fit the wrapped GP model which will induce posterior samples; $(\bftheta^*_b,\textbf{K}^*_b), b= 1,2,...,B$. Using Monte Carlo integration the following approximation is obtained:
	
	\begin{equation}
		E(e^{iX(s_0)}|\textbf{X}) \approx \frac{1}{B} \sum_{b}\exp(-\sigma^2(s_0,\bftheta^*_b)/2 + i\tilde{\mu}(s_0,\textbf{X}+2\pi \textbf{K}^*_b;\bftheta^*_b)).
	\end{equation}
	 The posterior mean kriged direction is 
	\begin{equation}
		\mu(s_0,\textbf{X})= \arctan^*(g_{0,s}(\textbf{X}),g_{0,c}(\textbf{X})),
	\end{equation}
	and the posterior kriged concentration is 
	\begin{equation}
		c(s_0,\textbf{X})=\sqrt{(g_c(s_0,\textbf{X}))^2 + (g_s(s_0,\textbf{X}))^2},
	\end{equation}
	which is induced if  $g_c(s_0,\textbf{X}) = \frac{1}{B} \sum_{b^*}\exp(-\sigma^2(s_0,\bftheta^*_b)/2) \cos(\tilde{\mu}(s_0, \textbf{X}+2\pi \textbf{K}^*_b;\bftheta^*_b))$ and
	\\
	$g_s(s_0,\textbf{X}) = \frac{1}{B} \sum_{b^*}\exp(-\sigma^2(s_0,\bftheta^*_b)/2) \sin(\tilde{\mu}(s_0, \textbf{X}+2\pi \textbf{K}^*_b;\bftheta^*_b))$.
	
\subsection{Projected Gaussian Spatial Process}

Suppose a random vector $\bm{Y}= (Y_1, ... Y_p)^{\prime}$ follows a $p$-dimensional multivariate Gaussian distribution, with mean $\bm{\mu}$ and covariance matrix $\bm{\Sigma} (p \ge 2)$. Then the unit vector $\bm{U}=\bm{Y}/||\bm{Y}||$ follows a projected Gaussian distribution with the same parameters and is denoted as $PN(\bm{\mu}, \bm{\Sigma})$ as defined in \cite{jupp2009directional}. When $p=2$, we obtain the circular projected Gaussian distribution. By projecting a bivariate spatial process on R$^2$, we can construct a spatial stochastic process of random variables taking values on a circle. Letting $(\cos\bm{X(s}), \sin\bm{X(s)})^{\prime} = (Y_1(s),Y_2(s))^{\prime}/||\bm{Y(s)}||$, we obtain the circular process $\bm{X(s)}$. This projected process inherits properties of the inline (linear) bivariate process such as stationarity. If we let $\bm{Y(s)}$ be a bivariate GP with mean $\bm{\mu(s)}$ and cross-covariance function $C(s,s^{\prime}) = \text{cov}(\bm{Y(s)},\bm{Y(s^{\prime})})$. Then the induced circular process upon projection is defined as the projected Gaussian spatial process (PGSP). For the choice of the cross-covariance function we let $C(s,s^{\prime}) = \zeta(s,s^{\prime})\cdot T$ where $\zeta$ is a valid correlation function and $T = \left(\begin{matrix}\tau^2 & \rho\tau\\ \rho\tau & 1\end{matrix}\right)$ is a $2 \times 2$ positive definite matrix as defined in \cite{ley2018applied}. 

Similarly to the WGSP, a Bayesian modelling framework is proposed for kriging due to the complexity of the conditional distributions of the GP. For the Bayesian formulation, we consider a conjugate prior, the bivariate Gaussian prior for $\bm\mu$. For $\tau^2$ an inverse gamma with mean = 1 and for $\rho$ a uniform prior $(-1, 1)$ is considered. For the decay parameter $\phi$ of the exponential correlation function, a
uniform prior with support allowing ranges larger than the maximum distance
over the region is utilised. The reader is referred to \cite{ley2018applied} for further details related to the projected Gaussian process.

\section{Results and Discussion}\label{sec:results}

The R (version 4.2.3 (2023-03-15 ucrt) \cite{R}) software package \textbf{CircSpaceTime} developed by \cite{jona2020circspacetime} was used for the modelling of the wind direction data in South Africa, specifically the \textbf{WrapSp}, \textbf{ProjSp}, \textbf{WrapKrigSp} and \textbf{ProjKrigSp} functions. \textbf{CircSpaceTime} was specifically developed for the implementation of Bayesian models for spatial interpolation of directional data using the wrapped Gaussian distribution and the projected Gaussian distribution.

Firstly, the \textbf{WrapSp} function was applied to estimate the wrapped Gaussian posterior distribution for the given wind data. The \textbf{WrapSp} function can run for multiple MCMC chains, storing the posterior samples for $\mu$ (circular mean), $\sigma^2$ (variance) and $\phi$ (spatial correlation decay parameter). Based on the data described in section \ref{sec:data} there were $97$ observations ($n = 97$), $87$ of these observations were used for the modelling while the other $10$ observations were our validation set. The validation set, consisting of $10$ randomly selected points from the $97$ observations, were used for prediction and model diagnostics. The \textbf{WrapSp} function requires the specification of prior distributions and a few parameters for the MCMC computation. The prior distribution values were chosen based on the data exploration as discussed in Section \ref{sec:data} and Table \ref{tab:descriptives}.

An exponential covariance function was considered. The prior for $\mu$ was a wrapped Gaussian distribution, for $\sigma^2 $ an informative inverse gamma prior and for the decay parameter $\phi$ a uniform  prior which is weakly informative were  considered. 
The details of the model specification were  provided for the 23:00 time period only. The remaining time periods follow similarly.
Therefore, the prior distribution values applied for the 23:00 time period data were :

\begin{itemize}
	\item $\mu \sim \text{WN}(0, 2)$
    \item $\sigma^2 \sim \text{IG}(7, 0.5)$
	\item $\phi \sim \text{U}(0.001, 0.9)$
\end{itemize} 

\begin{figure}[h!]
     \centering
     \begin{subfigure}[b]{0.49\textwidth}
         \centering
         \includegraphics[width=\textwidth]{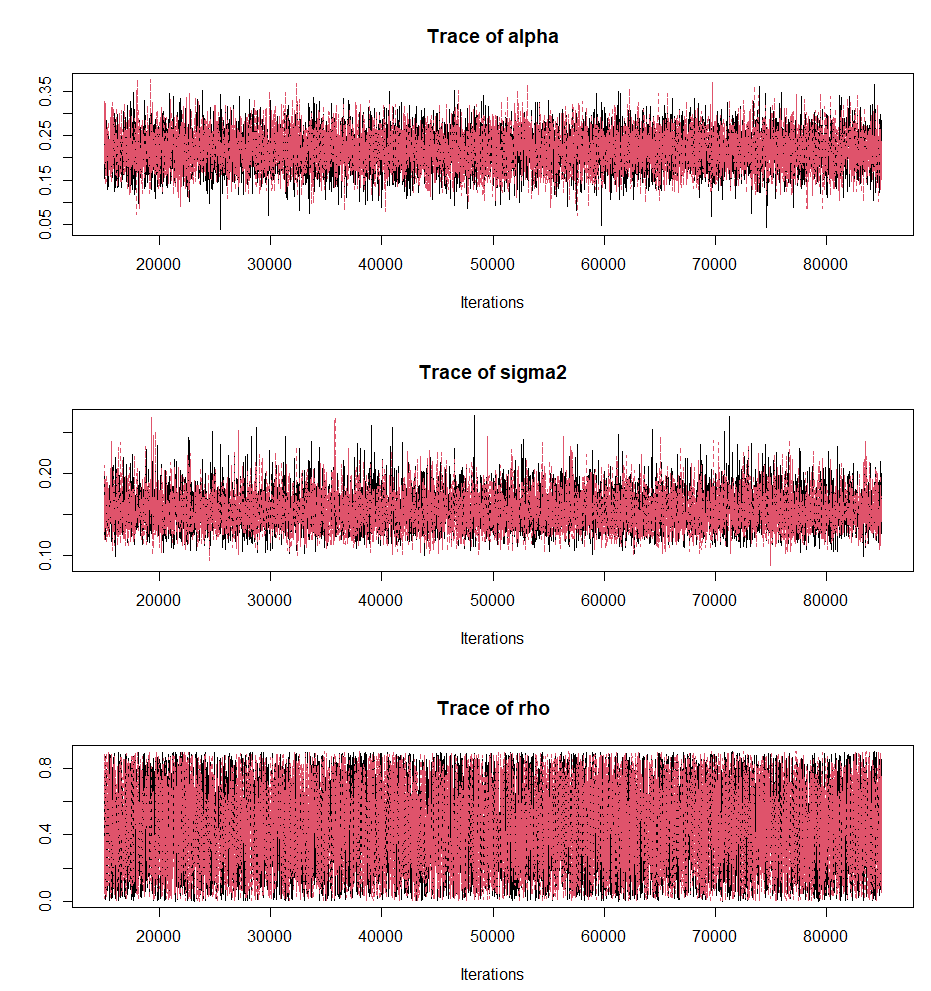}
         \caption{Trace plots of MCMC run for $\mu$ (top), $\sigma^2$ (middle) and $\phi$ (bottom).\label{fig6}}
     \end{subfigure}
     \hfill
     \begin{subfigure}[b]{0.49\textwidth}
         \centering
         \includegraphics[width=\textwidth]{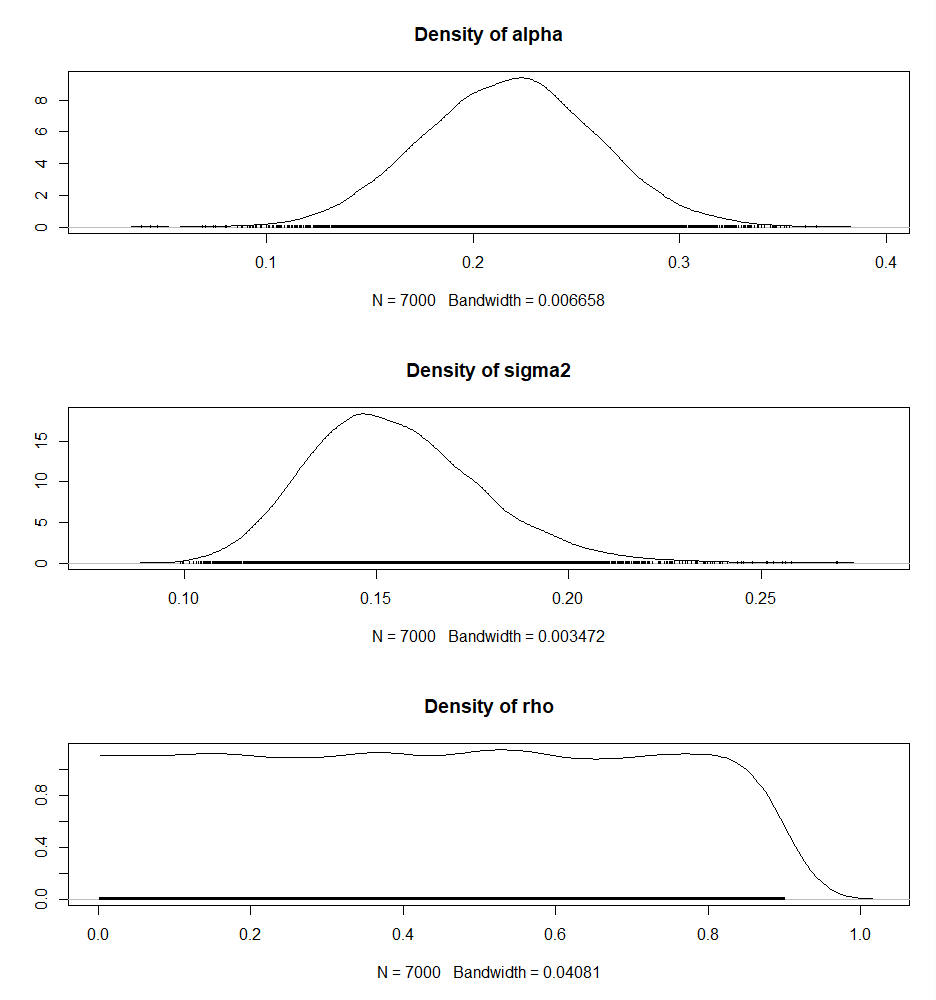}
         \caption{Density plots of MCMC run for $\mu$ (top), $\sigma^2$ (middle) and $\phi$ (bottom).\label{fig7}}
     \end{subfigure}
        \caption{Traces and densities from the MCMC run for the wrapped Gaussian spatial model.}
        \label{fig: MCMC runs}
\end{figure}

The MCMC ran with two chains in parallel for $100000$ iterations with a burnin of $30000$, thinning of $10$ and an acceptance probability of $0.234$ following \cite{jona2012spatial}. The adaptive process of the Metropolis Hasting step starts at the $100^{th}$ iteration and ends at the $10000^{th}$ iteration, it is important that the adaptive procedure end before the burnin is initiated to guarantee that the saved samples were  drawn from correct posterior distributions as in \cite{jona2020circspacetime}. The \textbf{ConvCheck} function was used to check for convergence and to obtain graphs of the MCMC. Figure \ref{fig: MCMC runs} illustrates the traces and densities of the MCMC. A traceplot is an essential plot for evaluating convergence and diagnosing chain problems. It shows the time series of the sampling process and the expected outcome is to get a traceplot that looks completely random. The traceplots and the estimated posterior density plots of the generated samples, are shown in Figure \ref{fig: MCMC runs} for each of the parameters. Using our fitted model the \textbf{WrapKrigSp} was applied for the interpolation. The function produces posterior spatial predictions on the unobserved locations across all posterior samples, together with the mean and variance of the corresponding linear Gaussian process. Once the predictions were  obtained the average prediction error (APE) - defined as the average circular distance - and circular continuous ranked probability score (CRPS) were  computed for the model, see \cite{jona2012spatial} and \cite{jona2020circspacetime}. 

\begin{figure}[h!]
	\begin{subfigure}{0.5\textwidth}
		\centering
		\includegraphics[width=\textwidth]{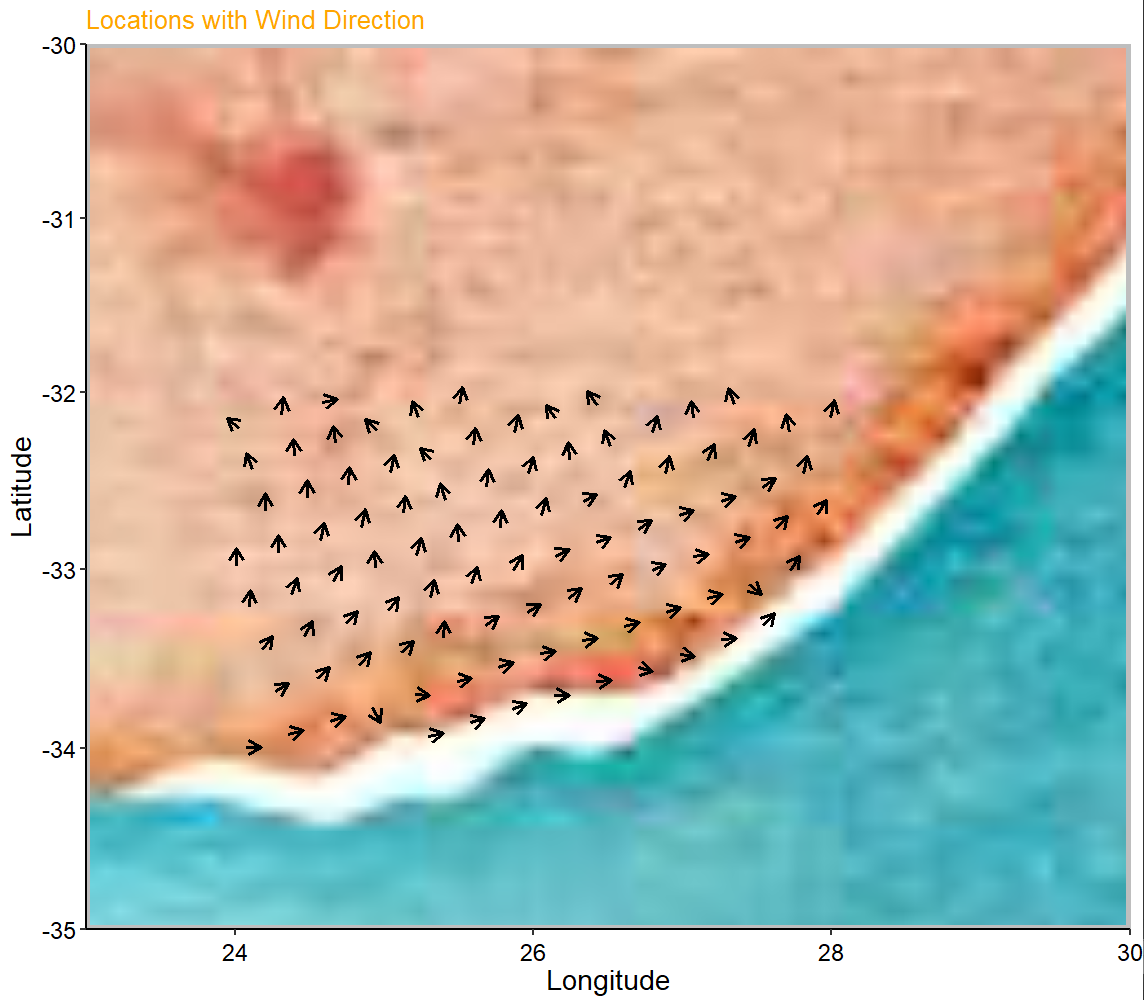}
		\caption{2012-12-31-05:00}
	\end{subfigure}
	\hfill
	\begin{subfigure}{0.5\textwidth}
		\centering
		\includegraphics[width=\textwidth]{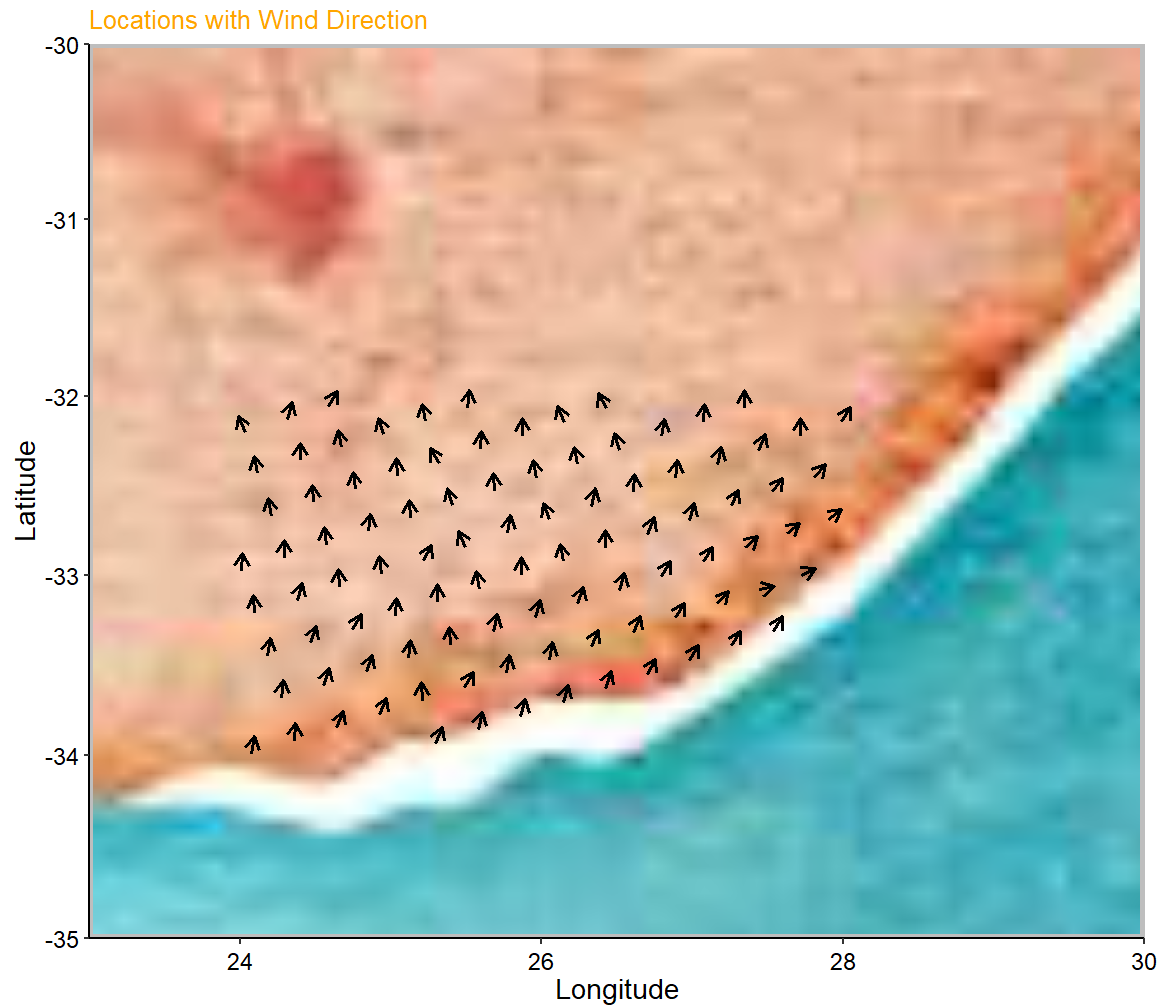}
		\caption{2012-12-31-23:00}
	\end{subfigure}
\caption{South Africa: observed wind directions over the considered region at different time periods.}
\label{directions}
\end{figure}

\begin{table}[h]
\caption{The $95\%$ credible intervals for $\hat{\mu}, \sigma^2, \phi$, the APE and CRPS for the WGSP model for the different time periods.}
\label{res}
\resizebox{\textwidth}{!}{
	\begin{tabular}{llllll}
		 & $\hat{\mu}$   & $\sigma^2$   & $\phi$      &   &    \\
		Time of day & $95\%$ C.I. & $95\%$ C.I.  & $95\%$ C.I.  &  APE & CRPS  \\ \hline
        05:00 & $(0.45406; 0.76958)$   & $(0.40963; 0.74301)$  & $(0.02375; 0.87666)$ & $0.66563$ & $0.45200$   \\ \hline
        11:00 & $(0.41816; 0.62425)$   & $(0.17859; 0.31959)$   & $(0.01624; 0.58459)$ & $0.58396$ & $0.47646$ \\ \hline
        17:00 & $(0.07741; 0.22503)$ & $(0.09163; 0.15986)$ & $(0.01686; 0.58482)$ & $0.12439$ & $0.06647$ \\ \hline
		23:00 & $(0.13639; 0.30191)$ & $(0.11557; 0.20323)$ & $(0.02382; 0.87776)$ & $0.21914$ & $0.14669$ \\ \hline
	\end{tabular}}

\end{table}

From Tables \ref{res} we observe $95\%$ credible intervals for $\hat{\mu}, \sigma^2, \phi$ as well as the APE and CRPS for the wrapped Gaussian model. One must take into account that the $\hat{\mu}$ is a directional variable. The APE scores can be attributed to the fact that only $97$ observations were considered. The APE score demonstrates sensitivity to the selection of prediction points, resulting in variability when different coordinates were  used in the validation set. 
The APE is very dependent on the number of observations considered and the prior selection for $\phi$. These results align with the conclusion in \cite{riha2020hyperpriorsensitivity} which emphasises the importance of hyper-parameter settings for the prior distributions of the spatial decay parameter $\phi$ and the variance $\sigma^2$ for spatial interpolation with wrapped Gaussian process models. We note that the APE is affected by the data's variability. As depicted in Figure \ref{directions} and observed in Table \ref{tab:descriptives} and \ref{res}, there is a noticeable contrast in data variance between the time periods at 05:00 and 23:00. Specifically, at 05:00, wind directions exhibit significant variability, whereas at 23:00, they tend to align in a more consistent direction. Consequently, this disparity in data variability contributes to the difference in APE scores between these time periods. A similar pattern emerges when comparing the conditions at 11:00 and 17:00. It can be noted that the two morning time periods have much more variability than the two evening time points, with 17:00 having the lowest variance of $0.05948$ yielding the lowest APE of $0.12439$ as well.

Next we fit the PGSP model to the wind data observed at the 23:00 time period. Note the PGSP is more sensitive to the choice of priors, specifically for the decay parameter. The details of the model specification were  provided for the 23:00 time period only. The remaining time periods follow similarly. The prior distribution values used in the PGSP model were :

\begin{itemize}
	\item $\bm\mu \sim \text{N}\left(\left(\begin{matrix}0 \\ 1 \end{matrix}\right), \left(\begin{matrix}10 & 0\\ 0 & 10\end{matrix}\right)\right)$
	\item $\sigma^2 \sim \text{IG}(7, 0.5)$
 	\item $\phi \sim \text{U}(0.001, 0.9)$
     \item  $\tau \sim \text{U}(-1,1)$
\end{itemize} 

We specify an exponential covariance function to be used. The prior for $\mu$ was a bivariate Gaussian distribution, for $\sigma^2 $ an informative inverse gamma prior, for $\tau$ a uniform prior and for the decay parameter $\phi$ a uniform  prior which is weakly informative were  considered. The remainder of the function specification was the same as the WGSP model. From the convergence check (including the traceplots that were  not reported) we see that the chains reached convergence. The PGSP's flexibility allows a better fit of the model with an APE of $0.03874$ and a CRPS of $0.02506$ for the 23:00 time period.

\begin{table}[h]
\centering
\caption{Goodness of fit measures for the wrapped Gaussian model (WGSP) and projected Gaussian model (PGSP) for the South African wind data.}
\label{tab:results compare}
\begin{tabular}{llll}\hline
\textbf{Time of day}            & \textbf{Model} & \textbf{APE} & \textbf{CRPS} \\ \hline
\multirow{2}{*}{\textbf{05:00}} & WGSP           &   $0.66563$           &    $0.45200$           \\
                                & PGSP            & $0.12156$   &  $0.09955$   \\ \hline
\multirow{2}{*}{\textbf{11:00}} & WGSP           &  $0.58396$        &   $0.47646$            \\
                                & PGSP            &  $0.06419$  & $0.04361$    \\ \hline
\multirow{2}{*}{\textbf{17:00}} & WGSP           &   $0.12439$           &    $0.06647$           \\
                                & PGSP            &  $0.04839$  &   $0.04319$  \\ \hline
\multirow{2}{*}{\textbf{23:00}} & WGSP           &   $0.21914$  &   $0.14669$ \\
                                & PGSP            &   $0.03874$  &  $0.02506$ \\ \hline
\end{tabular}
\end{table}

Table \ref{tab:results compare} reports the results of the APE and CRPS for both the WGSP and PGSP  models. It is clear that for the South African wind data the PGSP model outperforms and is able to better capture the structure of the data. The WGSP model was computationally less demanding and allowed the choice of less informative priors, as well as the parameters can be easily interpreted which is not the case for the projected Gaussian model. 

\section{Conclusion}\label{sec:con}
This paper explored the potential of utilising directional statistics within spatial analysis to model wind patterns in South Africa, drawing on methods developed in \cite{jona2012spatial}. The wrapped Gaussian model and projected Gaussian model were considered to account for the cyclic nature of the wind directions whilst also accounting for the spatial dependence. Based on the APE and CRPS, we conclude that the projected Gaussian process is an effective and precise approach to modelling wind patterns in South Africa. The model can adeptly manage directional data indexed by space, capturing the spatial structure among these observations. Looking ahead, enhancements to this model can be made through a more refined selection of parameters, like prior distributions, and by incorporating a more extensive set of locations to represent a broader area. Additionally, there is  potential to expand this model into a spatio-temporal model, accounting for time as well. Another avenue for future work resides in accounting for the wind speed (and other wind characteristics) along with the wind directions. 

In closing, the application of directional Gaussian processes in tandem with the capabilities of the \textbf{CircSpaceTime} package in \textbf{R} presents a compelling avenue for enhancing the accuracy and reliability of wind direction modelling. As the world increasingly recognises the critical role of sustainable energy sources, such as wind power, refining our understanding of wind behaviour becomes paramount, especially in South Africa with our current electricity problem. Better and more accurate understanding of wind behaviour can improve the design and optimisation of wind farms. Thus, ensuring efficient and effective harnessing of wind energy.

\bibliographystyle{unsrt}  
\bibliography{references}

\end{document}